\documentclass[twocolumn,showpacs,preprintnumbers,amsmath,amssymb]{revtex4}

\usepackage{graphicx}
\usepackage{dcolumn}
\usepackage{bm}

\begin{document}

\preprint{A7.02.094}

\title{A Bright, Guided Molecular Beam With Hydrodynamic Enhancement}

\author{David Patterson}
\author{John M. Doyle}%
 \email{doyle@physics.harvard.edu}
\affiliation{%
Department of Physics, Harvard University, Cambridge, Massachusetts
02138, USA \\ Harvard-MIT Center for Ultracold Atoms, Cambridge,
Massachusetts 02138, USA
}%

\pacs{39.10.+j Atomic and molecular beam sources and techniques}
\date{\today}

\begin{abstract}
We realize a novel high flux source of cold atoms and molecules
employing hydrodynamic enhancement of an effusive aperture 
at cryogenic temperatures. Molecular oxygen from the source is coupled
to a magnetic guide, delivering a cold, continuous, guided flux of
$3 \times 10^{12}$ O$_2$ s$^{-1}$.
The dynamics of the source are studied by creating and
spectroscopically analyzing high-flux beams of atomic ytterbium.
\end{abstract}

\maketitle

\section{\label{sec:level1}Introduction}

 The study of cold gas-phase molecules is a major thrust of modern chemical and atomic
 physics\cite{beamreview}.  In particular, inelastic collisions of cold molecules
have recently attracted widespread theoretical and experimental
interest. These collisions are a window into the internal structure
and dynamics of molecular interactions
 as well as being the expected fundamental limiting process for collisional cooling
 of molecules into and through the ultracold regime \cite{bohn01}. In addition to
 collision studies, cold
 molecules are sought as ideal test objects for precise measurement of fundamental constants
 (including the search for an electron EDM\cite{diatomicparity,hst02}) and as bits for
 quantum computing\cite{demille:067901,molensemblepreprint}. All of these areas of
 research benefit from high flux sources of cold molecules and this has spurred
 widespread interest in realizing such sources.

Although cold atomic sources have been laboratory standards for over
a decade, only recently have similar sources of molecules been
demonstrated. This is largely because the complex level structure of
molecules makes laser cooling -- the workhorse of atom cooling --
extremely difficult. Demonstrated techniques to produce cold, low
velocity samples of molecules are buffer-gas
cooling\cite{weinstein1998}, billiard-like
collisions\cite{chandler}, filtering slow molecules from an effusive
source \cite{elecfilter,nda03}, deceleration of molecules from a
seeded supersonic source in time-varying electric
fields\cite{beamreview,FirstStark}, photoassociation of cold and
ultracold atoms \cite{sage2005}, and the formation of molecules from
ultracold atoms using Feshbach resonances \cite{firstfeshbach}. In
general, the fluxes achieved with these molecular sources have been
far lower than those achieved in cold atom sources.

 We earlier\cite{maxwell2005} reported a molecular and atomic beam
source based on buffer gas cooling, showing that cryogenic cooling
of ablated atoms and molecules could effectively produce cold,
near-effusive beams. By employing a cold cell ($\approx 4$ K) filled
with helium gas and several solid ablation targets (e.g. PbO),
pulsed beams were produced with temperature and forward velocity
equivalent to 5 K. This was realized by producing gas phase
molecules with short-pulse laser ablation inside the cell. A
millimeter size orifice in the side of the cell allowed a small
fraction of the cold molecules to escape into a cryogenic
high-vacuum region, forming a beam. The ratio of the number of
molecules cooled in the cell to the number emitted into the beam was
determined by the ratio of the orifice area to the total area of the
interior of the cell, about $10^{-3}$. It was estimated that beams
of molecules from this source coupled to a magnetic or electric
guide would produce a peak guided flux no higher than $10^{9}$
molecules s$^{-1}$.

In this letter we realize a cold molecule buffer-gas source that
uses hydrodynamic effects to vastly enhance the molecular beam flux
compared to the old source. Oxygen is used as our test molecule. We
show that this new source can successfully load a magnetic guide,
producing a guided beam of cold O$_{2}$ with a continuous flux of $3
\times 10^{12}$ molecules s$^{-1}$ into a room-temperature
high-vacuum region. This is 1000 times greater than that predicted
with the old source. This flux compares favorably to cold
\emph{atom} sources and far exceeds previous demonstrated fluxes of
cold molecules. Using atomic ytterbium we further characterize the
creation and dynamics of the beam before it enters the guide. We
develop a simple model of hydrodynamic enhancement within the cell
and find reasonable agreement between this model and our data. The
performance of this guide/source combination is well described by
Monte-Carlo simulation.

\section{\label{apparatus}Beam Apparatus}
The heart of our experimental apparatus is a $\approx$ 2.5 cm size
cell anchored to the cold plate of a cryostat. This cell is a copper
box with two fill lines (one for helium and one for oxygen) on one
side, an aperture on the opposite side, a Yb ablation target inside,
and windows for laser access. In these experiments we create beams
of either atomic Yb or O$_{2}$. To produce beams of O$_{2}$ we flow
He and O$_{2}$ continuously into the cell where they mix and
thermalize to the temperature of the cell.  The He does not freeze
to the walls of the cell but any O$_{2}$ molecule that touches the
wall does. The O$_{2}$ cools by thermalizing with the helium. The
helium and a fraction of the O$_{2}$ escape out through an
double-stage aperture (see section \ref{sec:twostage}), forming a
beam, with the balance of the O$_{2}$ remaining frozen on the cell
walls. The same general processes of thermalization and beam
creation occur with Yb. To produce beams of Yb we flow He
continuously into the cell and ablate the Yb target with a pulsed
laser (4 nsec, 532 nm, 10 mJ).

\begin{figure*}
\includegraphics{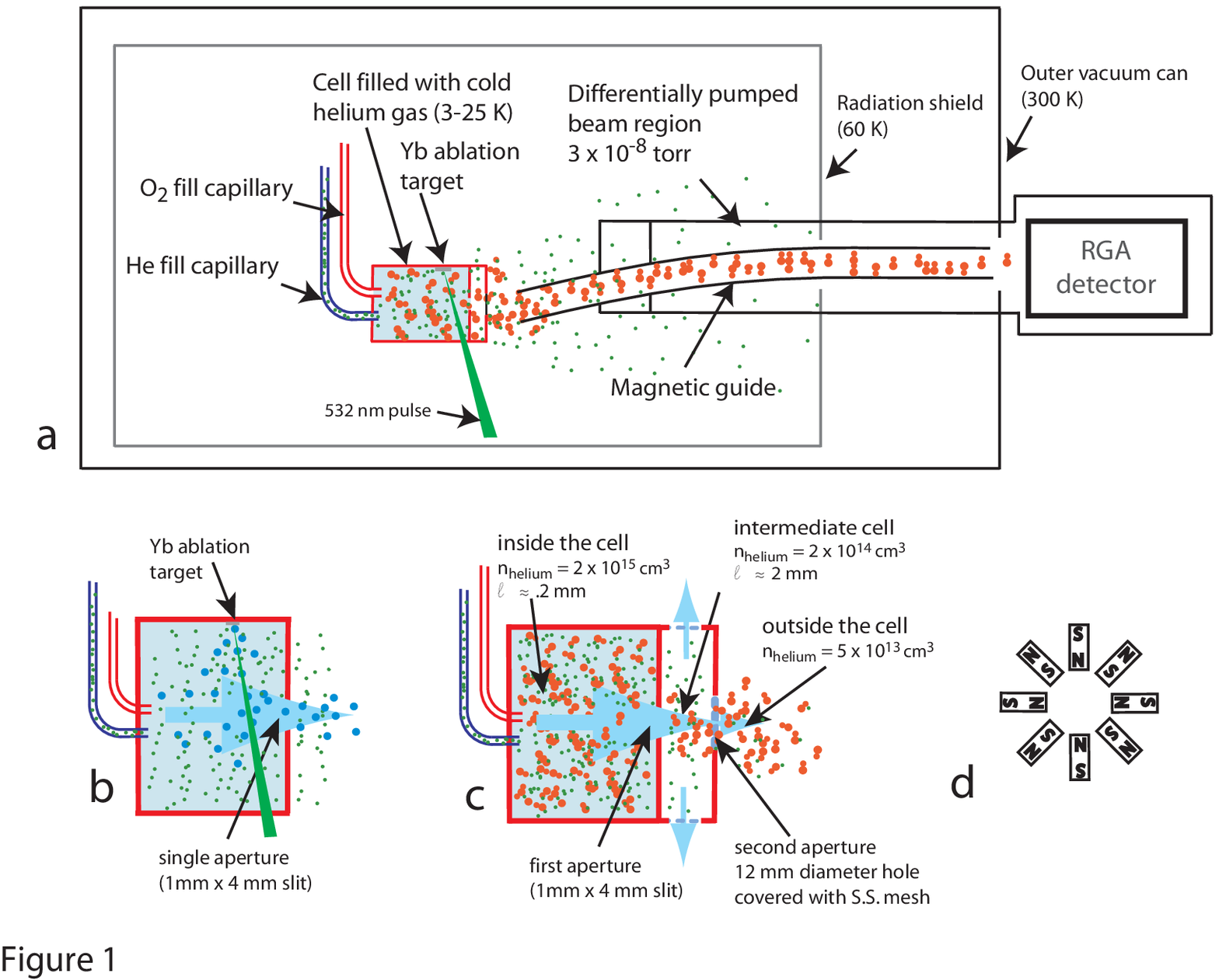}
\caption{ (color online) Experimental apparatus.\\ a) Hot ($\approx$
100K) O$_2$ molecules or ablated Yb atoms thermalize with helium gas
in a cold
  (3-25K) copper cell.  The mixture then sprays through the slit aperture and
  enters a beam guide.  Cold, low field seeking molecules are guided
  into an RGA detector.  \\b)``One-stage aperture'' detail.  This
  aperture was characterized with laser-ablated atomic Yb.
  The
  mean free path in the main cell (left) is generally short enough to
  thermalize hot Yb atoms before they touch and stick to the
  wall of the cell.  The slit aperture allows this mixture to form a beam in
  the dewar vacuum.  Hydrodynamic entrainment greatly enhances the flow through this
  aperture compared
  to a simple effusive source. This configuration produced high flux
  atomic Yb beams with significant forward velocity (see
  Fig.~\ref{ybdata} and discussion on
  page~\pageref{simplesetup}). \\c)``Two-stage aperture'' detail.  This two-stage aperture is
  implemented with the idea to produce a high flux effusive source.
  As in b), a mixture of He and O$_2$ first exits from the cell through a slit aperture.  However, in this case, rather than enter
  the beam directly, the mixture enters a small second cell. Additional venting holes
  on the side of the second cell maintain a low enough pressure that flow through the final
  aperture is largely effusive.
   The He density in the second chamber is much lower than
  in the first cell, so O$_2$ molecules  experience just a few more collisions before
  being diffusing through the final aperture.
  These collisions are sufficient to
  rethermalize the molecules that have been hydrodynamically boosted due to their
  passage through the first aperture.\\d) Magnet
  detail.  The magnetic guide is an octopole, composed of 8 NdFeB
  permanent magnets, and has a depth of $\approx$ 5000 gauss (680 mK
  for O$_2$).  Low field seekers are confined to the center of the
  guide.
} \label{trappic}
\end{figure*}

In more detail, the copper cell is $2.5 \times 2.5 \times 2.5 ~{\rm
cm}$ ($V \approx$ 15 cm$^{3}$) and its temperature is varied between
$T = $ 2.6 K - 25 K. It is surrounded by (dewar) vacuum, heat
shields and optical windows (see Fig.~\ref{trappic}a). Cooling of
the cell is done with a pulsed tube refrigerator and a small pumped
liquid helium reservoir (used to achieve temperatures below 4 K).
The copper He gas fill line to the cell is thermally anchored to a
1.3 K cold plate a few centimeters before it enters the cell.  The
O$_{2}$ fill line is held at 100 K, above the freezing point of
oxygen. A short stainless steel sleeve thermally disconnects the
relatively warm O$_{2}$ line from the cell. Helium is typically
flowed into the cell at $1 \times 10^{17}$- $8 \times 10^{18}$ atom
s$^{-1}$ and O$_{2}$ is flowed into the cell at similar rates,
typically $1 \times 10^{17}$- $3 \times 10^{18}$ molecules s$^{-1}$.

Despite the large He gas flow, the beam region vacuum is maintained
at a low $3 \times 10^{-8}$ torr by two stages of differential
pumping with high speed cryopumps made of activated charcoal ($T$ =
6.5K, 2000 cm$^2$ total apparent area). A 30 cm long magnetic guide
(see Fig.~\ref{trappic}d) can be placed with the guide entrance 1 cm
from the aperture. The guide is constructed from an array of
permanent NdFeB magnets (each magnet 0.3 cm $\times$ 0.6 cm $\times$
7.5 cm). The magnets are arranged in a linear octopole, such that
low field seeking molecules are confined to the center of the guide.
The far end of the guide (the ``guide exit'') is not at cryogenic
temperatures, but rather deposits the guided molecules into a room
temperature residual gas analyzer (RGA).  The guide has an inner
diameter of 0.9 cm and 5000 G depth, equivalent to 680 mK for O$_2$
in its most magnetic guidable state. The guide undergoes a single
bend with radius of curvature $r$ = 20 cm. This bend ensures that
only guided molecules can pass; i.e. O$_2$ must be moving slower
than 7000 cm s$^{-1}$ (9 K equivalent temperature) to be guided
around the bend. The entire system imposes a heat load of $\approx$
1 watt on the 4 K refrigerator.

\section{\label{results}Results}

We run two different types of experiments, guided (with O$_2$) and
unguided (with Yb). In the guided work, we beam O$_2$ into a
magnetic guide and measure  with the RGA the flux of O$_2$ exiting
the guide. The RGA is calibrated by comparison to a NIST-traceable
standard O$_2$ leak. In the unguided work, we remove the guide
assembly and use Yb to characterize the beam source. The flux and
velocity profile of the Yb beam and the density and temperature of
Yb gas in the cell are all measured using laser absorbtion
spectroscopy.

We chose to guide O$_2$ because it is relatively
easy to introduce into our cold cell and has a 
high magnetic moment of 2 Bohr magneton. Because the O$_{2}$ is
cooled in zero magnetic field, it is expected that the molecules
occupy a uniform statistical mixture of Zeeman states. Thus,
$\approx$~1/9 of the molecules exiting the aperture are in the
maximally guided $N = 1$,$J = 2$,$m_J = 2$ state, where $N$ is the
rotational quantum number, J is the total angular momentum quantum
number, and $m_J$ is its projection on the axis defined by the magnetic field.
We estimate that the contribution to guided beam
flux from other states is small ($<$ 10\%).

\begin{figure}
  \includegraphics{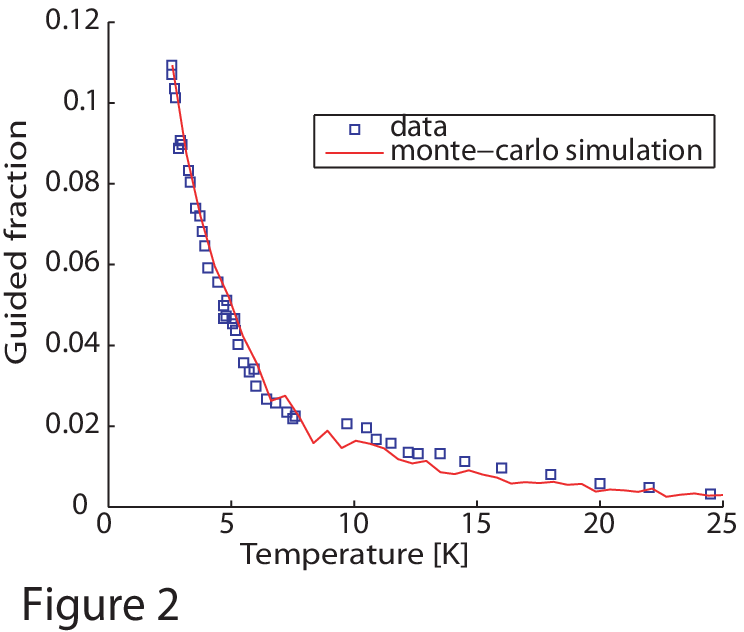}\\
  \caption{(color online) Guided flux of O$_2$
   as a function of cell temperature.  The temperature
  dependence of the measured flux is in excellent agreement with the Monte-Carlo
  simulation (solid line).  The vertical axis represents the absolute guided fraction
  (the ratio of the number of guided O$_2$ molecules to the number exiting the source)
  as calculated by the simulation.  The only free parameter in this fit is the
  absolute scale.
    In order to run for extended periods of time while avoiding
    plugging the aperture, this data was
taken with oxygen input (0.5 sccm) and lower helium input (0.4 sccm)
about one order of magnitude lower than that used to demonstrate our
peak flux of $3 \times 10^{12}$ O$_2$ s$^{-1}$. The highest flux
shown here represents $2.5 \times 10^{11}$ O$_2$ s$^{-1}$}
  \label{basicdata}
\end{figure}

 The molecule source and guide were characterized by varying the O$_{2}$ flow, He
flow, and cell temperature, while always monitoring the output flux
of O$_2$. Fig.~\ref{basicdata} shows our guided flux for a range of
temperatures. Because the depth of the guide (680 mK) is less than
the temperature of the source over the entire range of temperatures
used, a colder source will lead to a higher flux of guided
molecules. This is seen in the data. A Monte-Carlo simulation of our
guide is plotted with the data. Agreement is excellent, showing not
only that we are guiding molecules but that the details of the
system are well understood.

Our maximum measured guided flux of $3 \times 10^{12}$ O$_2$
s$^{-1}$ is achieved by simultaneously flowing 10 sccm of O$_2$ and
2 sccm of He into the cell (sccm is standard atmosphere-cc per
minute). The system can operate at this peak flux for about 100
seconds before the aperture becomes plugged with oxygen ice.

We note that we are confident that the O$_2$ in our guide is cold
rotationally as well as translationally.  In all our previous work
with buffer gas cooling and in many supersonic beam experiments it
has been observed that rotational thermalization rates are
comparable or faster than elastic scattering
rates\cite{weinstein1998,supersonics}.

\subsection{Cell and Aperture Dynamics}

The high guided fluxes of O$_2$ we achieve is due to a newly
utilized effect in our beam source, hydrodynamic enhancement. We
first describe a simple theory of this effect and then describe
experiments with Yb that verify this basic picture.

The helium gas in our cell can be characterized by mean free path
$\ell$ (typically,  $\ell \approx$ 0.1 cm). Atoms and molecules can
leave the cell via two processes: by diffusing to the cell wall
(with a small fraction of the molecules diffusing out of the 
aperture) or by being entrained in the helium flow that is
continuously leaving through the 
aperture. If the diffusion
time for an atom or molecule ($\tau_{diffusion}$) is much shorter than the
characteristic time a helium atom (or entrained species) spends
in the cell ($\tau_{pumpout}$), hydrodynamic effects can be ignored.
In this diffusive limit the fraction that escape into the
beam from a cell of side length $L$ with aperture area $A$
is $f \approx {\pi L^2}/{A}$ (about about $2 \times 10^{-3}$ for our
cell). If, on the other hand, $\tau_{diffusion} \gg \tau_{pumpout}$,
a large fraction are entrained inside the cell and are
forced to move toward the 
aperture and eventually into the
beam. The dimensionless parameter
  \begin{equation}
\gamma \equiv \frac{\tau_{diffusion}}{\tau_{pumpout}}
\end{equation}
therefore provides a good indication of which limit the cell is in,
with $\gamma \ll 1$ means diffusion limit and $\gamma \gg 1$ means
entrainment limit. We note that the entrainment being discussed here
is \emph{not} the same effect that is seen in typical supersonic
nozzles, where collisions \emph{within the nozzle} (as opposed to
the volume well before the nozzle) are important and flow velocities
are large compared to the sonic velocity. The mean flow velocity of
He within our cell is typically much less than the speed of sound in
the He gas.

 Exact analysis depends on geometrical
details, but it can be easily shown that  $\gamma = {kA}/{4 \ell
L}$, with $k$ a dimensionless constant of order unity. Ideally one
wishes to operate with as large a value of $\gamma$ (large helium
density) as possible while maintaining an effusive source. If the
density of helium just outside the aperture is large enough that the
mean free path is smaller than the aperture size, then entrainment
outside the aperture results in a significant increase in the mean
forward velocity (i.e. ``forward boost'') of the beam.  This is a
detriment to guiding experiments as the boost makes a smaller
fraction of the total beam flux guidable.

Experimental study of these processes is needed to verify our model.
As O$_2$ is very difficult to detect optically, we chose to
characterize the flux and velocity distribution of the beam with
atomic Ytterbium, which has easily accessible optical transitions.

\begin{figure}
  \includegraphics{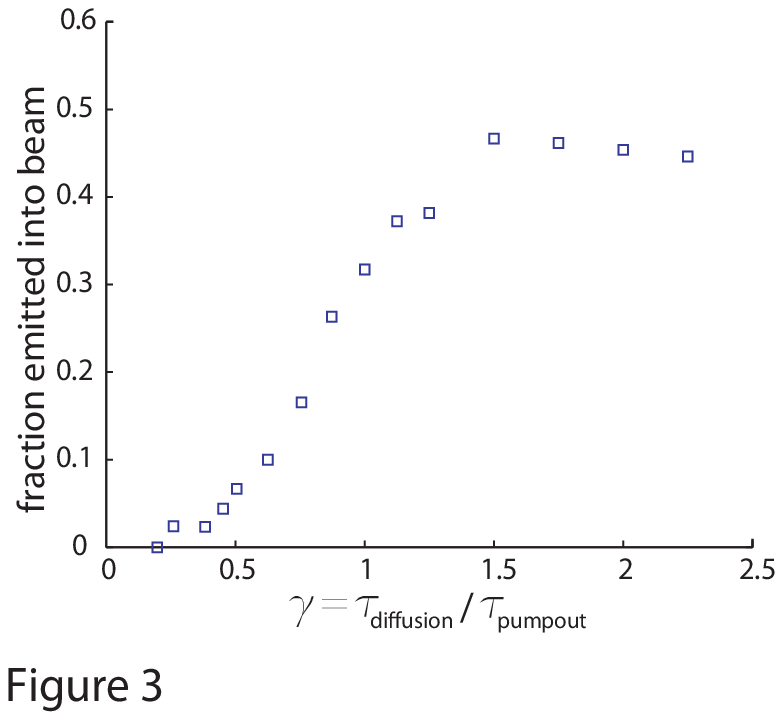}\\
  \caption{Plotted is the ratio of the number of ablated and cooled ytterbium atoms
  to the number emitted into the beam for a single stage aperture. At high buffer
  gas densities, which correspond to long diffusion times and therefore
  high $\gamma$ (see main text), 
  up to 40\% of the cold ablated atoms are
  detected in the 
  beam. The peak instantaneous flux
  represented here is $5 \times 10^{15}$ atom s$^{-1}$ sr$^{-1}$
  }
  \label{ybdata}
\end{figure}

Two aperture configurations were demonstrated.  Fig.~\ref{ybdata}\label{simplesetup}
shows the output efficiency of a ``one-stage aperture'' (see
Fig.~\ref{trappic}b) consisting of a simple slit aperture (1 mm x 4
mm), where this efficiency is defined as the ratio of the number of
Yb in the beam to the number of cold Yb produced by ablation in
the cell. At buffer gas flows where $\gamma
> 1$ (high flows), up to 40\% of the cold Yb atoms in the cell are
detected in the resultant beam. The divergence of the beam is
measured by comparing the doppler shifts in the absorbtion spectra
parallel and transverse to the atomic beam. The beam is somewhat
collimated compared to a pure effusive source, with a divergence of
$\approx$ 0.1 steradian. The peak flux of $6
\times 10^{15}$ atom s$^{-1}$ sr$^{-1}$ ($5 \times 10^{12}$
atoms/pulse), suggests that this simple cryogenic one-stage
aperture provides an attractive alternative to seeded supersonic
jet sources of cold molecules\cite{hindsjetsource}.  For trapping,
where total kinetic energy must be less than the trap depth,
typically a few K, this beam's high forward boost (velocity = 130 m s$^{-1}
\approx 160 ~ K$ effective temperature) makes it unsuitable.

Trapping work would ideally use a beam with both high flux and
minimal forward boost. In an attempt to achieve this best of both
worlds, the ``two-stage aperture'' shown in Fig.~\ref{trappic}c was
developed.\label{sec:twostage}
In this two stage aperture the He-Yb mixture still passes through
the 1 mm x 4 mm slit with a large hydrodynamic flux enhancement and
corresponding boost.  But, instead of directly entering the vacuum
beam region, this boosted beam enters a small chamber.
Atoms can pass from the small chamber through a 12 mm diameter aperture on the opposite side.
This second, larger aperture is covered by a
stainless steel mesh with pore size 140 microns (28 percent
transparency) and creates a near-effusive beam in the vacuum region.
The He density in the second chamber would be far too low to thermalize hot
atoms entering the cell. However, it is high enough to thermalize the cold (but
boosted) Yb that comes through the first orifice and low enough so that $\gamma \ll 1$, i.e.
diffusion dominated dynamics. The final
aperture is large enough to pass into the beam about 4\% of atoms that arrive in
the second chamber.  Finally, it is hypothesized that
the small pores in the mesh may help reduce hydrodynamic,
non-effusive flow inside and just outside the nozzle. Preliminary results in our lab with and without the mesh
provide some evidence that this is the case.

The Yb beam from the two-stage aperture has a mean velocity 35
m s$^{-1}$, with a spread of 20 m s$^{-1}$. The total beam flux of
$5 \times 10^{10}$ atoms for each Yb ablation pulse represents about
1\% of the cold atomic Yb produced in the ablation, or about 3\% of
the output of the one-stage aperture. In our guided O$_{2}$ experiments, the two-stage aperture is used.

\section{Potential Applications}

  The one-stage aperture cold source described above
   can provide high fluxes of any atom or molecule.
For example, under the assumption of $10^{12}$ cooled radicals of CaF per ablation pulse
\cite{egorov2001} and the use of a 15 Hz pulsed YAG laser, a 15\% duty cycle pulsed
  beam would be produced with a continuous flux of $7 \times 10^{13}$
  molecules s$^{-1}$ sr$^{-1}$ (10 ms  pulse length, 0.1 sr divergence,
  peak flux $5 \times 10^{14}$ molecules s$^{-1}$sr$^{-1}$).
  These fluxes are 500-1000 times higher than
  typical fluxes from seeded supersonic beams of radicals at
  comparable translational and rotational
  temperatures\cite{hindsjetsource}. As described above, we 
  produce even brighter cold atomic beams, with
a continuous flux of $7 \times 10^{14}$
  atom s$^{-1}$ sr$^{-1}$ (peak flux $5 \times 10^{15}$
  atom s$^{-1}$ sr$^{-1}$).
  This source is therefore an attractive option for collisional studies
  and high precision measurements based on cold
  molecules as well as atoms 
  \cite{hst02,rareearthparity}.

The two-stage aperture source could be applied to trapping polar
molecules. Observing collisions in trapped samples
  of polar molecules has proved to be difficult, primarily because available fluxes of cold polar
  molecules are typically low.  Replacing the magnetic guide
  demonstrated here with a simple electrostatic guide (as in
  \cite{rjr03,bethlem:491}) could realize a guided source of polar molecules with a
  phase space density 300 times higher than has been demonstrated to date.
  We note that the trap lifetimes for polar molecules in
  a 300 K environment are limited to a few seconds by optical pumping via thermal microwaves,
  a constraint which is
  conveniently avoided by our cryogenic
  environment\cite{opticalpumpingblackbody}.

Although in these experiments atoms and molecules were introduced
into the cell using ablation and gaseous flow, direct loading from a
molecular beam (e.g. oven or pulsed discharge) is also
possible\cite{egorov2004,newnh}. Thus, this source is truly general,
being able to cool molecules produced by any technology.

\section{Conclusion}
  We have demonstrated production of cold, guided O$_2$ with a flux
  of
   $3 \times 10^{12}$ O$_2$ s$^{-1}$.  We have further
  demonstrated a general long-pulse source of cold, unguided atoms or molecules with
  peak flux of
  $5 \times 10^{14}$ Yb s$^{-1}$.  The efficiency of the unguided source can be near unity,
  with the number atoms or molecules in the cold beam being up to 40\% of the number cooled
  in the cell.
\section{Acknowledgements}
  We are grateful for the contributions of Y. Takahashi and group
  members by introducing us to Yb for use in beam experiments. This
  work was funded by NSF Grant PHY-0457047.

\bibliography{final}

\end{document}